\documentclass[preprintnumbers,superscriptaddress,amsmath,amssymb,prd,preprint,nofootinbib]{revtex4}
\usepackage{amssymb}
\usepackage{mathrsfs}
\usepackage{amsthm}
\usepackage{graphicx}
\usepackage{amsfonts}
\usepackage[colorlinks,linkcolor=red,anchorcolor=blue,citecolor=green]{hyperref}
\usepackage{subfigure}
\usepackage{float}
\usepackage{enumerate}
\usepackage{multirow}

\usepackage{array}
\usepackage[figuresright]{rotating}
\usepackage{upgreek}

\begin{document}
	
	\title{Latest data constraint of some parameterized dark energy models}
	\author{Jing Yang}
	\affiliation{Division of Mathematica and Theoretical Physics, Shanghai Normal University, 100 Guilin Road, Shanghai 200234, China}

	\author{Xin-Yan Fan}
	\affiliation{Division of Mathematica and Theoretical Physics, Shanghai Normal University, 100 Guilin Road, Shanghai 200234, China}
	
	\author{Chao-Jun Feng}
	\email[]{fengcj@shnu.edu.cn}
	\affiliation{Division of Mathematica and Theoretical Physics, Shanghai Normal University, 100 Guilin Road, Shanghai 200234, China}
	
	\author{Xiang-Hua Zhai}
	\email[]{zhaixh@shnu.edu.cn}
	\affiliation{Division of Mathematica and Theoretical Physics, Shanghai Normal University, 100 Guilin Road, Shanghai 200234, China}	
	
\begin{abstract}
	 Using various latest cosmological datasets including Type-Ia supernovae, cosmic microwave background radiation, baryon acoustic oscillations, and estimations of the Hubble parameter, we test some dark energy models with parameterized equations of state and try to distinguish or select observation-preferred models. We obtain the best fitting results of the six models and calculate their values of the Akaike Information Criteria and Bayes Information Criterion. And we can distinguish these dark energy models from each other by using these two information criterions. However, the $\Lambda $CDM model remains the best fit model. Furthermore, we perform geometric diagnostics including statefinder and Om diagnostics to understand the geometric behaviour of the dark energy models. We find that the six DE models can be distinguished from each other and from  $\Lambda$CDM, Chaplygin gas, quintessence models after the statefinder and Om diagnostics were performed. Finally, we consider the growth factor of the dark energy models with comparison to $\Lambda $CDM model. Still, we find the models can be distinguished from each other and from $\Lambda $CDM model through the growth factor approximation.

\end{abstract}

\maketitle
	
	\section{Introduction}{\label{Sec.1}}
	
	At the end of the 20th century, the observations of Type-Ia supernovae (SNIa) first indicated that the universe is under accelerating expansion \cite{SupernovaCosmologyProject:1998vns,SupernovaSearchTeam:1998fmf}. Later, various experimental observations, including the large-scale structure (LSS)\cite{SDSS:2003eyi,SDSS:2004kqt} and the  cosmic microwave background radiation (CMB)\cite{WMAP:2003elm,WMAP:2003ivt,WMAP:2006bqn,Planck:2011aj,Planck:2011ijl,Planck:2011qep}, also provided evidence for the accelerating expansion of the universe. In order to explain the acceleration of the universe, the efforts that continue to date include two aspects. On the one hand, the gravitational part may be modified and extended theories of gravity may be constructed  \cite{Nojiri:2006ri,DeFelice:2010aj,Capozziello:2011et,Cai:2015emx}. On the other hand, a matter component with negative pressure, i.e. dark energy (DE), which may drive the acceleration, is introduced into the matter part (see review articles on DE Ref.\cite{Carroll:2000fy,Peebles:2002gy,Bartelmann:2009te,Sahni:2004ai,Copeland:2006wr,Li:2011sd,Li:2012dt,Frieman:2008sn,Straumann:2006tv}). The currently preferable and also the simplest cosmological model is the $\Lambda$ cold dark matter ($\Lambda$CDM) model, in which the cosmological constant $\Lambda$ plays the role of DE. However, due to the deficiencies of the $\Lambda$CDM model such as the cosmic coincidence issue and fine-tuning problem\cite{Weinberg:2000yb,Guo:2004vg,Guo:2004xx,Guo:2007zk,Yang:2015tzc}, the dynamical DE models have been widely discussed\cite{Ratra:1987rm,Zlatev:1998tr,Brax:1999yv,Barreiro:1999zs,Garriga:1999vw,Caldwell:1999ew,Caldwell:2003vq,Sola:2005et,Roy:2022fif,Yang:2018qmz}. 
	
	Due to the fact that there is no preferable DE model that can completely describe the dynamical phenomena of the universe, several attempts have been made in recent years to model them from observations. Parameterization of cosmological or DE parameters is one of the most concerned attempts. One can parameterize the Hubble parameter, the deceleration parameter or the energy density of DE \cite{Ozer:1985wr,Abdel-Rahman:1990eaa,Vishwakarma:1998un,Vishwakarma:2001ac,Abdussattar:2011zza,Pacif:2014uaa,Mamon:2017jom,Rezaei:2019xwo,Roy:2022fif}. Parameterizing the DE density parameter, for example using a simple power law expansion $\Omega_{DE}=\sum_{i=0}^{N}A_{i}z^{i}$, where $z$ is the red shift, is also a frequently used method\cite{Alam:2003fg,Alam:2004jy,Daly:2003iy,Daly:2004gf,Gong:2004ns,Jonsson:2004zk,Alam:2004ip}. Following this approach, one can parameterize the equation of state (EoS) of the DE as $w(z)=\sum_{i=0}^{N}w_{i}z^{i}$\cite{Weller:2000pf,Huterer:2000mj,Weller:2001gf,Astier:2000as}, which is a simple parameterization of the EoS that can describe the dynamical evolutionary behavior for a large number of DE models. However, this parameterization diverges at high redshifts. Furthermore, since the angular diameter distance depends on the form of $w(z)$ and the angular scale features of the CMB temperature anisotropy varies with the peak, the constraint on the angular diamete distance to the last scattering surface by the CMB would be problematic. Then, a stable parameterization of the EoS $w(z)=w_{0}+w_{1}z/(1+z)$ that extends the parameterization of the DE to redshifts $z\gg 1$  was given and has been widely discussed \cite{Chevallier:2000qy,Linder:2002et,Choudhury:2003tj,Feng:2004ad,Gong:2004sd,Gong:2005de,Yang:2018prh,Escamilla:2021uoj,Jassal:2004ej}. Furthermore, a modified model $w(z)=w_{0}+w_{1}z/(1+z)^{2}$ was proposed by Jassal et al.\cite{Jassal:2004ej}. It can model a DE component that has the same value of EoS at the present time and at high redshifts. Both models are bounded at high redshifts $z\gg 1$, but they cannot be distinguished. In Ref.\cite{Gong:2005de}, Gong et.al. also  proposed two one-parameter models and the model $w(z)=w_{0}{\rm e}^{z/(1+z)}/(1+z)$ such that $w=0$ in the future $z \to -1$. Feng et.al.\cite{Feng:2012gf} proposed two-parameter models $w\left(z\right)=w_{0}+\frac{w_{1}z}{1+z^{2}}$ and $w\left(z\right)=w_{0}+\frac{w_{1}z^{2}}{1+z^{2}}$, with $w\left(z\right)$ bounded in the future for both models. In Ref.\cite{Perkovic:2020mph}, the authors classified the parameterized EoS models proposed in recent years by the number of parameters and gave the range of model parameters using numerical analysis. In the appendix of Ref.\cite {Pacif:2020hai}, the authors made a summary of various parameterized models in recent years. In our work, we aim to explore the cosmological feasibility of some DE models with parameterized EoS using recent observations to constrain the models.
	
	A large number of DE models may produce similar evolutionary behaviors and, correspondingly, similar histories of cosmic expansion. Therefore, effective differentiation of models is very important. Using various data to test DE models, to select good models, or to compare models, has become a standard approach in cosmological studies. However, by using only one kind of observational data to constrain models, a degeneracy of certain cosmological parameter among different models usually occurs. In order to break this degeneracy, joint constraints of multiple observational data are often used. In our work, we will use the combination of supernova data, the temperature and polarization anisotropy data from the CMB, the baryon acoustic oscillations (BAO), and Hubble parameter observational H(z) data. For supernova data, we will compare the constraints from two samples: Joint Light-curve Analysis (JLA) and Pantheon, where the redshift range is extended in the latter.

	In this paper, we will use the above mentioned observational data to investigate the cosmological feasibility of six DE models with parameterized EoS \cite{Linder:2002et,Gong:2005de,Jassal:2004ej,Feng:2012gf}, and analyze the DE nature with geometric diagnostics. The paper is organized as follows: In Sec.\ref{Sec.2}, six parametric DE models are reviewed. Section \ref{Sec.3} gives the constraints of the models with observational data. In Sec.\ref{Sec.4} we perform two diagnostic analyses that can distinguish the models, and analyze the impact of DE on the matter density perturbation by the growth factors of the models. We conclude our study in the last section.

	\section{The considered six parameterized dark energy models}{\label{Sec.2}}
	
	According to Einstein's gravitational field equation and the flat Fridmann-Robertson-Walker (FRW) metric, the Friedman equations are
	\begin{equation}\label{Fre}
		\begin{array}{c}
			3H^{2}=\rho_{m}+\rho_{r}+\rho_{de},\\
			3H^{2}+2\dot{H}=-(p_{m}+p_{r}+p_{de}), 
		\end{array}
	\end{equation}
	where $H=\frac{\dot{a}}{a}$ is the Hubble parameter, $a$ is the scale factor, and the dot is the derivative with respect to cosmic time. $\rho_i$ and $p_i$ are the energy density and pressure, in which $i=m,r$ and $de$ denote matter, radiation and DE, respectively.  Here, we use units $8\pi G=1$.
	
	Assuming that there is no interaction between dark matter and DE in the universe, the equations of energy conservation are,
	\begin{equation}\label{energy}
		\begin{array}{c}	
			\dot{\rho_{de}} +3H \left(1+w\right)\rho_{de}=0,\\
			\dot{\rho_{m}} +3H \rho_{m}=0,	
		\end{array}
	\end{equation}
   and it is assumed that the matter in the universe is dusty matter with $p_{m}$=0.
	
	Firstly, we give a review of the six parameterized DE models that we will consider in the following sections.
	
	Model 1: It was proposed by Gong and Zhang \cite{Gong:2005de} with the EoS,
	\begin{equation}\label{model1}
		w\left(z\right)=\frac{w_{0}}{1+z},
	\end{equation}
	where the only parameter $w_{0}$ is constant. For this model, $w\left(z\to 0\right)= w_{0}$ is the current value of the EoS. The model is bounded by $w\sim 0$ at high redshift $z\gg 1 $, which means at that time DE is represented as dust matter. But in the future $w\left(z\to-1\right)\sim \infty$, this model will have singularity. Combining Eqs.(\ref{Fre}), (\ref{energy}), and  (\ref{model1}), one obtains
	\begin{equation}\label{model1 e}
		E^{2}\left(z\right)=\left(1-\Omega_{m0}-\Omega_{r0}\right)\left(1+z\right)^{3}  {\rm e}^{\frac{3w_{0}z}{1+z}}+\Omega_{m0}\left(1+z\right)^{3}+\Omega_{r0}\left(1+z\right)^{4},
	\end{equation}
	where $E=\frac{H}{H_{0}}$ is the dimensionless Hubble parameter, $H_{0}$ is the current value of the Hubble constant. $\Omega_{i0}=\frac{\rho_{i0}}{3H_0^2} $ are the current values of density parameters. 
	
	Model 2: It is also an one-parameter model proposed by Gong and Zhang\cite{Gong:2005de}, and its EoS is
	\begin{equation}\label{model2}
		w\left(z\right)=\frac{w_{0}}{1+z}{\rm e}^{\frac{z}{1+z}}.
	\end{equation}
	Compared with Model 1, the EoS in the future $z\to-1$ for Model 2 is $w\to 0$ where DE represents dust matter. Similarly, one can obtain
	\begin{equation}\label{model2 e}
			E^{2}\left(z\right)=\left(1-\Omega_{m0}-\Omega_{r0}\right)\left(1+z\right)^{3}{\rm e}^{3w_{0}\left({\rm e}^{\frac{z}{1+z}}-1\right)}+\Omega_{m0}\left(1+z\right)^{3}+\Omega_{r0}\left(1+z\right)^{4}.
	\end{equation}
	
	 Model 3: For two-parameter models, the parameterized EoS can be expressed as
	 \begin{equation}
	 	w\left(z\right)=\frac{p}{\rho}=w_{0}+w_{1}f\left(z\right),
	 \end{equation}
	 where $w_{0}$, $w_{1}$ are constants and $f\left(z\right)$ is a function of the redshift $z$. Different forms of $f\left(z\right)$ correspond to different DE models. Also, the model will return to $\Lambda $CDM model when $w_{0}=-1$ and $w_{1}=0$. 
	 
	 The model with $f(z)=\frac{z}{1+z}$ is known as the Chevallier-Polarski-Linear(CPL) model\cite{Linder:2002et},
	\begin{equation}\label{cpl}
		w\left(z\right)=w_{0}+\frac{w_{1}z}{1+z}.	
	\end{equation}
	This model is divergent when describing future evolution. Similarly, one has
	\begin{equation}\label{cpl e}
		E^{2}\left(z\right)=\left(1-\Omega_{m0}-\Omega_{r0}\right)\left(1+z\right)^{3\left(1+w_{0}+w_{1}\right)}{\rm e}^{\frac{-3w_{1}z}{1+z}}+\Omega_{m0}\left(1+z\right)^{3}+\Omega_{r0}\left(1+z\right)^{4}.	
	\end{equation}
	
	This is the Model 3 we will consider.
	
	Model 4: This model\cite{Jassal:2004ej} is a modification of Model 3 and its EoS is as follows,
	\begin{equation}
	w\left(z\right)=w_{0}+\frac{w_{1}z}{(1+z)^{2}},	
	\end{equation}
	which diverges in the future as Model 3. And we can obtain
	\begin{equation}\label{e4}
		E^{2}\left(z\right)=\left(1-\Omega_{m0}-\Omega_{r0}\right)\left(1+z\right)^{3\left(1+w_{0}\right)}{\rm e}^{\frac{3w_{1}z^{2}}{2(1+z)^{2}}}+\Omega_{m0}\left(1+z\right)^{3}+\Omega_{r0}\left(1+z\right)^{4}.	
	\end{equation}

	Model 5: In Ref.\cite{Feng:2012gf}, Feng et. al. proposed two-parameter models that are different from the CPL model and can describe the evolutionary behavior of the universe from the past to the future. Model 4 is one of the models with
	\begin{equation}\label{model4}
		w\left(z\right)=w_{0}+\frac{w_{1}z}{1+z^{2}}.	
	\end{equation}
	Then, 
	\begin{equation}\label{e5}
			E^{2}\left(z\right)=\left(1-\Omega_{m0}-\Omega_{r0}\right)\left(1+z\right)^{3\left(1+w_{0}-\frac{1}{2}w_{1}\right)}{\rm e}^{\frac{3}{2}w_{1}{\rm arctan}z}\left(1+z^{2}\right)^{\frac{3}{4}w_{1}}+\Omega_{m0}\left(1+z\right)^{3}+\Omega_{r0}\left(1+z\right)^{4}.	
	\end{equation}
	
	 Model 6: It is another two-parameter model in Ref.\cite{Feng:2012gf} with the EoS 
	\begin{equation}\label{model5}
		w\left(z\right)=w_{0}+\frac{w_{1}z^{2}}{1+z^{2}}.	
	\end{equation}
	Similarly, one has
	\begin{equation}
			E^{2}\left(z\right)=\left(1-\Omega_{m0}-\Omega_{r0}\right)\left(1+z\right)^{3\left(1+w_{0}+\frac{1}{2}w_{1}\right)}{\rm e}^{-\frac{3}{2}w_{1}{\rm arctan}z}\left(1+z^{2}\right)^{\frac{3}{4}w_{1}}+\Omega_{m0}\left(1+z\right)^{3}+\Omega_{r0}\left(1+z\right)^{4}.
	\end{equation}

	The behaviors of the EoSs of the above six models are listed in Tab \ref{1}. Among these models, the EoSs of Model 1, Model 3 and Model 4 will be divergent at $z\to -1$. However, the divergence can be avoided in Model 2, Model 5 and Model 6. Moreover, for Models 1 and 2 when $w_{0}=-1$, the EoSs are the same as that of $\Lambda$CDM at the present time. For Model 3, the EoS is the same as $\Lambda$CDM  in the past and at present time when $w_0=-1$ and $w_1=0$. The EoS of Model 4 is the same as $\Lambda$CDM in the past and at present time when $w_0=-1$. And Models 5 and 6 return to $\Lambda$CDM when $w_0=-1$ and $w_1=0$.
	
	\begin{table}
		\centering 
		\caption{Behaviors of EoS parameters for different DE models.}
		\label{1}
		\begin{tabular}{m{2.5cm}<{\centering}|m{3cm}<{\centering}|m{1cm}<{\centering}|m{1.5cm}<{\centering}|m{1.5cm}<{\centering}}
			\hline
			Mode l           &  $w\left(z\right)$                                                 & $z=0$             & $z\to\infty$                         &$z\to-1$                 \\
			\hline
			Model 1 \cite{Gong:2005de}   	&  $w\left(z\right)=\frac{w_{0}}{1+z}$                        & $w_{0}$           & $0$                   &$\infty$                  \\
			\hline
			Model 2 \cite{Gong:2005de}      &  $w\left(z\right)=\frac{w_{0}}{1+z}{\rm e}^{\frac{z}{1+z} }$        & $w_{0}$           & $0$                   &$0$                   \\
			\hline
			Model 3	\cite{Linder:2002et}        &  $w\left(z\right)=w_{0}+\frac{w_{1}z}{1+z}$                 & $w_{0}$           & $w_{0}+w_{1}$         &$\infty$                 \\
			\hline
			Model 4 \cite{Jassal:2004ej}     &$w\left(z\right)w_{0}+\frac{w_{1}z}{(1+z)^{2}}$   &$w_{0}$  &$w_{0}$  &$\infty$\\
			\hline
			Model 5 \cite{Feng:2012gf}       &  $w\left(z\right)=w_{0}+\frac{w_{1}z}{1+z^{2}}$             & $w_{0}$           & $w_{0}$               &$w_{0}-\frac{w_{1}}{2}$     \\ 
			\hline	
			Model 6 \cite{Feng:2012gf}     	&  $w\left(z\right)=w_{0}+\frac{w_{1}z^{2}}{1+z^{2}}$         & $w_{0}$           & $w_{0}+w_{1}$         &$w_{0}+\frac{w_{1}}{2}$      \\
			\hline                  
		\end{tabular}
	\end{table}

	\section{Data Constraints}{\label{Sec.3}}

	The fast development of observational techniques has led to increasingly refined observational data and gradually increased constraints on DE models. Moreover, in order to break down the degeneracy in cosmological parameters among different models, the combination of multiple observational data is frequently used to give subtle constraints. In this section, we constrain DE models mentioned in the previous section with the Pantheon+CMB+BAO+H(z) and the JLA+CMB+BAO+H(z) data sets. And the constraint results are used to distinguish or compare the models. A brief introduction to these datasets are given in the following.

	\subsection{Pantheon and JLA sample} 
	We use the Pantheon SNIa sample at redshifts of $0.01<z<2.3$ \cite{Pan-STARRS1:2017jku}, which is a combination of Pan-STARRS1 (PS1), Sloan Digital Sky Survey (SDSS), Supernova Legacy Survey (SNLS), some low redshift data, and the Hubble Space Telescope (HST) sample of a total of 1048 supernovae. And the JLA dataset contains a total of 740 SNIa at redshifts of $0.01<z<1$\cite{SDSS:2014iwm}, including several samples at low redshifts ($z < 0.1$), all three seasons of SDSS-II ($0.05 < z < 0.4$) and three years of SNLS ($0.2 < z < 1$).
	
	Due to the linear relationship between the extreme luminosity of SNIa bursts with the colour of the light and the rate of decrease of the light-change curve, they are treated as standard candles for the detection of cosmological distances in astronomical observations. The luminosity distance is expressed in astronomical observations by introducing a distance modulus $m_{cor}-M$\cite{Gao:2020irn,Xu:2021xbt}, i.e.
	\begin{equation}
		\mu=m_{cor}-M=5{\rm log}_{10}\left(d_{L}/{\rm Mpc}\right)+25,
	\end{equation}
	where $m_{cor}$ is the corrected magnitude in the Pantheon observable, $d_{L}=\left(1+z\right)\int_{0}^{z}\frac{dz^{'}}{H\left (z^{'}\right ) }=\left(1+z\right)r\left(z\right)$ is the photometric distance and $r\left(z\right)$ is the co-movement distance. Therefore, the actual observed distance of SNIa modulus is
	\begin{equation}
		\mu_{obs}=m_{B}-M+\alpha x_{1}-\beta c+\bigtriangleup_{M}+\bigtriangleup_{B}, 
	\end{equation}
	where $m_{B}$ is the apparent magnitude in the $B$-band, $M$ is the absolute magnitude of the SN in the $B$-band when the stretching parameter $x_{1}=0$ and the color parameter $c=0$, $\alpha$ is the luminosity versus stretching coefficient, $\beta$ is the luminosity versus color number, $\bigtriangleup_{M}$ is a distance correction based on the host galaxy mass of the SN, and $\bigtriangleup_{B}$ is a distance correction based on the deviation predicted from the simulation (for details, see Ref.\cite{Pan-STARRS1:2017jku}).
	The $\chi^{2}$ for the SNIa dataset is
	\begin{equation}
		\chi^{2}_{SNIa}=\left(\mu_{i}^{obs}-\mu_{i}^{th}\left(p\right)\right)\left(Cov_{SN}^{-1}\right)_{ij}\left(\mu_{j}^{obs}-\mu_{j}^{th}\left(p\right)\right),
	\end{equation}
	where $p=(p_{1},...,p_{n})$ is a vector of $n$ fit parameters, $Cov_{SN}$ is the covariance matrix describing the systematic error of the supernova observations, $\mu_{i}^{obs}$ and $\mu_{i}^{th} $ are the observed and theoretical distance moduli, respectively.

	\subsection{Cosmic Microwave
		Background (CMB)}
	Generally, three observables from CMB observations are used to constrain the cosmological model parameters: the redshift $z_{*}=1048\left[1+0.00124\left ( \Omega _{b}h^{2}\right)^{-0.738} \right]\left [ 1+g_{1}\left ( \Omega _{m}h^{2} \right )^{g_{2}}  \right ]  $ during the photon decoupling period, the position of the first peak in the temperature rise power spectrum $l_{A}=\left(1+z_{*}\right)\frac{\pi D_{A}z_{*}}{r_{s}\left (z_{*}\right ) } $, and the translation parameter $\Re \equiv \sqrt{\Omega _{m}H_{0}^{2}} \left ( 1+z_{*} \right ) D_{A}\left ( z_{*} \right )$.
	
	The $\chi^{2}$ of the CMB is
	\begin{equation}
		\chi^{2}_{CMB}=\Delta p_{i}\left [ Cov_{CMB}^{-1 }\left ( p_{i},p_{j} \right ) \right ]\Delta p_{j}, 
	\end{equation}
	where $\Delta p_{i}=p_{i}-p_{i}^{data}$ , $p_{i}=\left\{l_{A},\Re,w_{b}\right\} $, and $w_{b}=\Omega_{b}h^{2}$.
	Using the Planck 2018 \cite{Planck:2019nip} data including temperature (TT), the polarization (EE), the cross-correlation of temperature and polarization (TE) power
	spectra, where the low multipoles (low-$ \ell $) temperature
	Commander likelihood in TT, the low-$ \ell $ SimAll
	likelihood in EE and the high multipoles (high-$ \ell $) part of Planck TT, TE, EE are contributed in the range $\ell \in \left [2,2500 \right ]$, one can obtain the central value of $p_{i}^{data}$ and the error in the $1 \sigma$ confidence interval (The relevant results are given in Ref.\cite{Planck:2019nip}).

	\subsection{Bayon Acoustic Oscillatons (BAO)}

	BAO are used in astronomical observations as a standard ruler for cosmological measurements of cosmic distances, and their data are analyzed from BAO eigenpower spectra or correlation functions in large scale surveys of galaxy clusters to extract the physical quantities: redshift and distance.
	
	In the direction of sight there is 
		\begin{equation}\label{H}
			H\left(z\right)=\frac{c*\delta z_{s}\left ( z \right )}{r_{s}\left(z_{d}\right)},
		\end{equation}
	where $r_{s}\left(z_{d}\right)$ is the co-moving acoustic horizon during the baryon towing period, and $\delta z_{s}$ denotes the BAO characteristic redshift distance along the direction of sight with redshift $z$.	
	
	In the direction perpendicular to the line of sight there is 
	\begin{equation}
		d_{A}\left(z\right)=\frac{r_{s}\left ( z_{d} \right )}{\theta_{s}\left(z\right)\left(1+z\right)},
	\end{equation}
	where $\theta_{s}\left(z\right)$ denotes the angle at which the BAO feature is opened perpendicular to the line of the sight direction. The data of BAO features can now be directly measured out of the value of the effective distance $D_{V}\left(z\right)$ corresponding to the redshift,
	\begin{equation}
		D_{V}\left(z\right)=\frac{r_{s}\left ( z_{d} \right )}{\left[\theta_{s}\left(z\right)^{2}\delta z_{s}
			\left(z\right)\right]^{\frac{1}{3}}}=\left[\frac{\left(1+z\right)^{2} d_{A}^{2}
			\left(z\right)cz}{H\left(z\right)}\right]^{\frac{1}{3}}.
	\end{equation}
	
	The $\chi^{2}$ of BAO is
	\begin{equation}
		\chi^{2}_{BAO}=\Delta p_{i}\left [ Cov_{BAO}^{-1 }\left ( p_{i},p_{j} \right ) \right ]\Delta p_{j}.	
	\end{equation}
	
	In our work, we utilize BAO measurements from 6dFGS \cite{Beutler:2011hx}, SDSS-MGS\cite{Ross:2014qpa} and BOSS DR12 \cite{BOSS:2016goe} as published by Planck 2018 results \cite{Planck:2019nip}.

	\subsection{H(z) Observational Data (OHD)}
	
	The Hubble parameter can characterize the evolution rate of the universe, and to some extent, the size and age of the universe. With the progress of the observational techniques, the accuracy of the current value of the Hubble constant has greatly improved. We use in our analysis a total of 57 data points\cite{Sharov:2018yvz} from direct observations of $H\left(z\right)$ by both the age differential measurement method and the BAO effect.
	
	The $\chi^{2}$ of the observed data for $H\left(z\right)$ is
	\begin{equation}
			\chi^{2}_{OHD}=\sum_{i=1}^{57}\frac{\left[H_{th}-H_{obs}\right]^{2}}{\sigma _{H\left({z_{i}}\right)}^{2}}.  
	\end{equation}
	
	The total  $\chi^{2}$ combining the above four datasets  is
	\begin{equation}
			\chi^{2}_{tot}= \chi^{2}_{SNIa}+\chi^{2}_{CMB}+\chi^{2}_{BAO}+\chi^{2}_{OHD}. 
	\end{equation}
	The minimum values of $\chi^{2}$, i. e., the $\chi^{2}_{min}$-values, reflect the goodness of fit for models. That is, the smaller $\chi^{2}_{min}$-values indicate that the model can be better supported by the current observation data. Since the $\chi^{2}_{min}$-values decrease as the number of the model parameters increases, it is unlikely to make a fair judgement on the merit of the model by the $\chi^{2}_{min}$-values alone. Considering the influence of the number of parameters, we will compare the models by using two information criterions, Akaike Information Criteria (AIC)\cite{Akaike} and Bayes Information Criterion (BIC)\cite{Schwarz:1978tpv},
	\begin{equation}
		{\rm AIC}=\chi^{2}_{min}+2K,
	\end{equation}
	\begin{equation}
		{\rm BIC}=\chi^{2}_{min}+K{\rm ln}N,
	\end{equation}
	where $K$ is the number of free parameters and $N$ is the total data points.
	
		We use the Markov Chain Monte Carlo (MCMC) method to explore the parameter space of these six parametrized DE models mentioned in Sec.\ref{Sec.2} and fit the data using the Pydm$\footnote{https://github.com/shfengcj/pydm}$ package written by our own group. In this section, we use two sets of combined datas, i.e., JLA+CMB+BAO+H(z) (JCBH) and Pantheon+CMB+BAO+H(z) (PCBH) data sets, to obtain the best-fit values of the parameters of different DE models and their 1$\sigma$ to 2$\sigma$ confidence levels. The results are shown in Tabs.\ref{fit1} and \ref{fit2}. One can see that the EoS parameter crosses -1 at a certain redshift in Model 3 and Model 6. Taking the best fit results from JCBH data for instance, in Model 3, the value of the EoS parameter at $z=0$ is $w_0=-0.9$, and it becomes $w_0+w_1=-1.41$ at $z=\infty$. And in Model 6, the value of the EoS parameter at $z=0$ is $w_0=-0.95$, and it becomes $w_0+w_1=-1.58$ at $z=\infty$. This crossing behavior of the EoS parameter cannot be realized in single field models such as quintessence models, but can be realized in two-field models (see Refs.\cite{Guo:2004fq,Guo:2006pc} for more details).
		
\begin{figure}[h]
		
		\centering
		\includegraphics[width=0.6\linewidth]{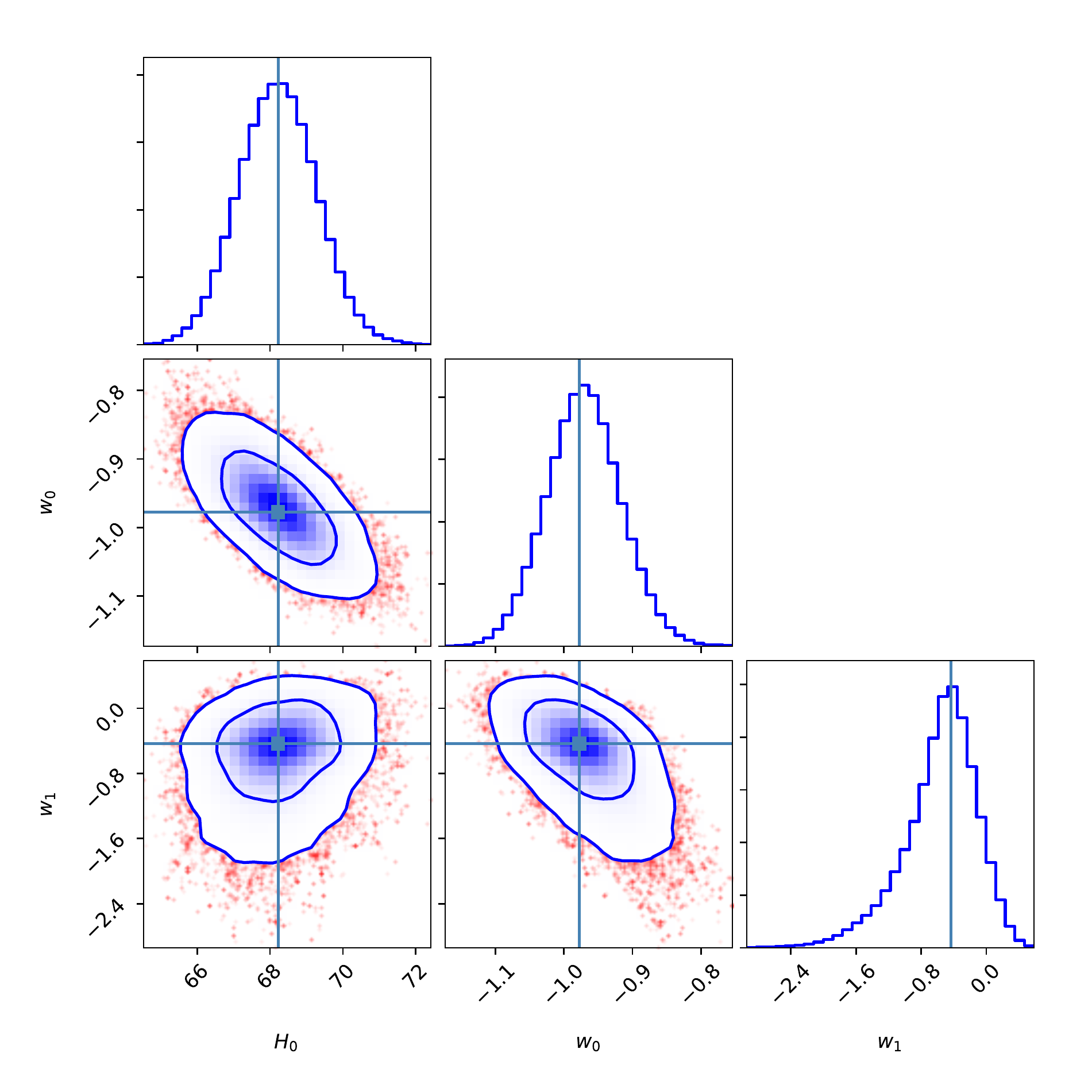}
		\caption{ Best-fit values of model parameters and their from 1$\sigma$ to 2$\sigma$ confidence levels for Model 6 by using PCBH data.}
		\label{fig:M5}
		
	\end{figure}

		 For better analysis and comparison, the $\Lambda$CDM model is chosen as a reference. The $\chi^{2}_{min}$-values reflect the goodness of fit, and we notice that Model 6 has the smallest $\chi^{2}_{min}$-value in Tabs.\ref{fit1} and \ref{fit2}, which is supported by current observational data. The best-fit values of model parameters for Model 6 and 1$\sigma$ to 2$\sigma$ confidence level are given in Fig.\ref{fig:M5} by using PCBH data. However, as we mentioned previously, the $\chi^{2}_{min}$-values are not suitable for comparing models because of the $\chi^{2}_{min}$-values are affected by the number of parameters. Here, we use the difference of AIC between a certain DE model and the $\Lambda$CDM model, $\bigtriangleup {\rm AIC}=\bigtriangleup \chi ^{2}_{min}+2\bigtriangleup K $, and the difference of BIC, $\bigtriangleup {\rm BIC}=\bigtriangleup \chi ^{2}_{min}+2\bigtriangleup K{\rm ln}N $, to quantitatively compare the models. We show in Tab.\ref{fit3}, the comparison of the models using the two information criterions. It is noted that for Models 1-6, $\bigtriangleup {\rm AIC}>2$ and $\bigtriangleup {\rm BIC}>2$, then, according to Refs.\cite{Kass:1995loi,Xu:2016grp,Rezaei:2019roe}, the best model is still the $\Lambda$CDM model.

		 The main difference between the two sets of combined datas is that the covariance matrices of the two supernova samples, Pantheon and JLA, are analyzed differently.  The covariance matrix in the JLA sample depends explicitly on the parameters $\alpha$ and $\beta$, so these two parameters must be added as additional parameters in the MCMC search. In contrast, for Pantheon sample, they do not need to be changed during the MCMC search because the effects of $\alpha$ and $\beta$ have been considered in the covariance calculation. As described in Ref.\cite{Jones:2017udy}, the constraints of PCBH and JCBH  are known to be in good agreement. We note that although the PCBH datasets cover a larger range of redshifts, it still has poor constraints for $w_{1}$ of Models 3-5. Meanwhile, the PCBH constraint results show that for Models 3-6 compared to JCBH, the values of $w_{0}$ are more negative and the values of $w_{1}$ are less negative, which is consistent with the result of Ref.\cite{DiValentino:2020evt}. For Models 1 and 2, the fit results prefer to choose less negative values of $w_{0}$.

	\begin{table}[thp]
			 
			\setlength{\abovecaptionskip}{0.1cm}
			\setlength{\belowcaptionskip}{0.9cm}
			\caption{The best fit values of the model parameters and 1$\sigma$ confidence level of Models 1,2 and $\Lambda$CDM model under two sets of combined data.}
	\label{fit1}
	\resizebox{\textwidth}{!}{
		\begin{tabular}{c|c|c|c|c|c|c}
			\hline
			\hline
			\multirow{2}*{Parameter} &\multicolumn{2}{c|}{$\Lambda$CDM} &\multicolumn{2}{c|}{Model 1}  &\multicolumn{2}{c}{Model 2} \\
			\cline{2-7}
			~&JCBH&PCBH &JCBH&PCBH &JCBH&PCBH\\
			\hline
			$H_{0}$  &$ 68.26^{+0.45}_{ -0.51}$ &$68.31^{+0.47}_{-0.50}$         &$70.97^{+1.16}_{-1.16}$&$69.86^{+0.99}_{-0.99}$ &$69.20^{+1.18}_{-1.09}$&$68.90^{+0.97}_{-0.98}$                        \\
			$w_{0}$  &$-$&$-$  &$-1.31^{+0.05}_{ -0.05}$&$-1.26^{+0.04}_{-0.05}$  &$-1.06^{+0.05}_{-0.05}$&$-1.04^{+0.04}_{-0.04}$                          \\
			\hline
			$\chi^{2}_{min}$  &$720.06$&$1062.87$    &$745.31$&$1090.19$ &$725.20$&$1067.81$                           \\
			\hline
			\hline

	\end{tabular}}
\end{table}

	\begin{table}[thp]
			 
			\setlength{\abovecaptionskip}{0.1cm}
			\setlength{\belowcaptionskip}{0.9cm}
			\caption{The best fit values of the model parameters and 1$\sigma$ confidence level of Models 3-6 under two sets of combined data.}
			\label{fit2}
			\resizebox{\textwidth}{!}{
		\begin{tabular}{c|c|c|c|c|c|c|c|c}
			\hline
			\hline
			\multirow{2}*{Parameter}  &\multicolumn{2}{c|}{Model 3} &\multicolumn{2}{c|}{Model 4} &\multicolumn{2}{c|}{Model 5} &\multicolumn{2}{c}{Model 6}  \\
			\cline{2-9}
			~&JCBH&PCBH  &JCBH&PCBH &JCBH&PCBH &JCBH&PCBH \\
			\hline
			$H_{0}$   &$67.94^{+1.30}_{-1.23}$&$68.24^{+1.01}_{-1.00}$    &$68.31^{+1.21}_{-1.32}$&$68.33^{+1.04}_{-1.02}$ &$68.17^{+1.32}_{-1.23}$&$68.37^{+1.08}_{-1.03}$ 
			&$67.84^{+1.27}_{-1.36}$&$68.21^{+1.03}_{-1.03}$                        \\
			$w_{0}$   &$-0.90^{+0.09}_{-0.09}$&$-0.95^{+0.09}_{-0.08}$  &$-0.93^{+0.11}_{-0.11}$&$-0.97^{+0.12}_{-0.11}$  &$-0.94^{+0.10}_{-0.10}$       &$-0.98^{+0.09}_{-0.08}$    &$-0.95^{+0.07}_{-0.07}$       &$-0.97^{+0.05}_{-0.05}$                        \\
			$w_{1}$  &$-0.51^{+0.33}_{-0.45}$&$-0.34^{+0.39}_{-0.45}$  &$-0.49^{+0.65}_{-0.76}$&$-0.26^{+0.75}_{-0.85}$
			&$ -0.29^{+0.33}_{-0.38}$&$-0.14^{+0.33}_{-0.39}$ &$-0.63^{+0.33}_{-0.44}$       &$-0.50^{+0.33}_{-0.49}$                        \\
			\hline
			$\chi^{2}_{min}$    &$719.07$&$1062.05$  &719.85&1062.60  &$719.49$&$1062.56$ &$ 718.18$&$1061.53$                         \\
			\hline
			\hline

	\end{tabular}}
\end{table}

\begin{table}[thp]
	
	\setlength{\abovecaptionskip}{0.1cm}
	\setlength{\belowcaptionskip}{0.9cm}
	\caption{The results of AIC and BIC of PCBH data for the models.}
	\label{fit3}
	\resizebox{\textwidth}{!}{
		\begin{tabular}{c|ccccccc}
			\hline
			\hline
			Criterion  &$\Lambda$CDM  & Model 1  &Model 2 & Model 3 &Model 4 &Model 5 &Model 6\\
			\hline
			AIC    &1066.87  &1096.16  &1073.81  &1070.05 &1070.60 &1070.56 &1069.53                \\
			\hline
			BIC    &1076.89 &1111.22 &1088.84  &1090.08 &1090.63 &1090.59 &1089.56          \\
			\hline
			\hline					
	\end{tabular}}
\end{table}

	\section{diagnostics of DE models}{\label{Sec.4}}
	
	In the previous section, we are able to distinguish DE models from each other with observational constraints. With the purpose of investigating the cosmological behaviour of various DE models and distinguishing one model from others, we will further use statafinder diagnostic, Om diagnostics, and growth factor to distinguish Models 1-6 from each other as well as from $\Lambda$CDM, Quintessence, and Chaplygin gas models.

	\subsection{Statefinder diagnostics}
	 Sahni et.al.\cite{Alam:2003sc,Sahni:2002fz} proposed a statefinder diagnostics $\left\{r, s\right\}$, which is a geometric-parametric method to distinguish  different DE models. The parameters are defined as follows	
	\begin{equation}\label{4.1}
		r=\frac{\dddot{a}}{aH^{3}},
	\end{equation}
	\begin{equation}\label{4.2}
		s=\frac{r-1}{3\left (q-\frac{1}{2}\right)},\left (q\ne \frac{1}{2}\right ).
	\end{equation}
	where $q=-\frac{\ddot{a}}{aH^{2}}$ is the deceleration parameter.
	
	Different $(r,s)$ pairs correspond to different DE models\cite{Alam:2003sc,Sahni:2002fz}. For example,
	\begin{itemize}
		\item  $(r=1,s=0)\to \Lambda $CDM model   \\
		\item  $(r<1,s>0)\to $Quintessence  model  \\
		\item  $(r>1,s<0)\to $Chaplygin gas (CG) model \\
	\end{itemize}

\begin{figure}[htbp]
		
		\centering
		\includegraphics[width=0.67\linewidth]{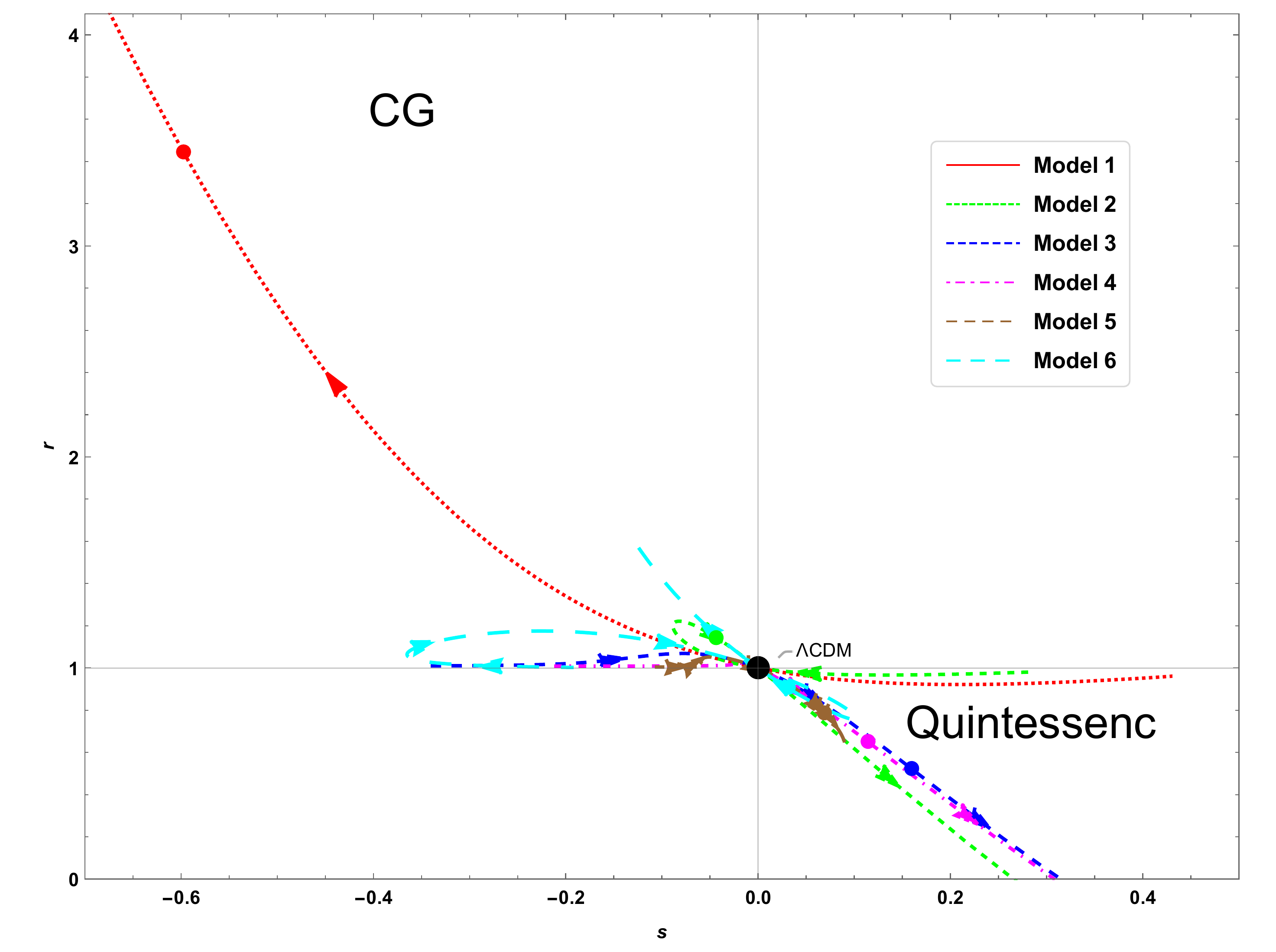}
		\caption{Evolution trajectories of six DE models in the $r-s$ plane, where the arrows indicate the temporal evolution of these models. The solid dots on these lines indicate the current states of the models and the point (0,1)  represents the $\Lambda $CDM model. }
		\label{fig:S-R}
	
	\end{figure}

\begin{figure}[htbp]
	
	\centering
	\includegraphics[width=0.67\linewidth]{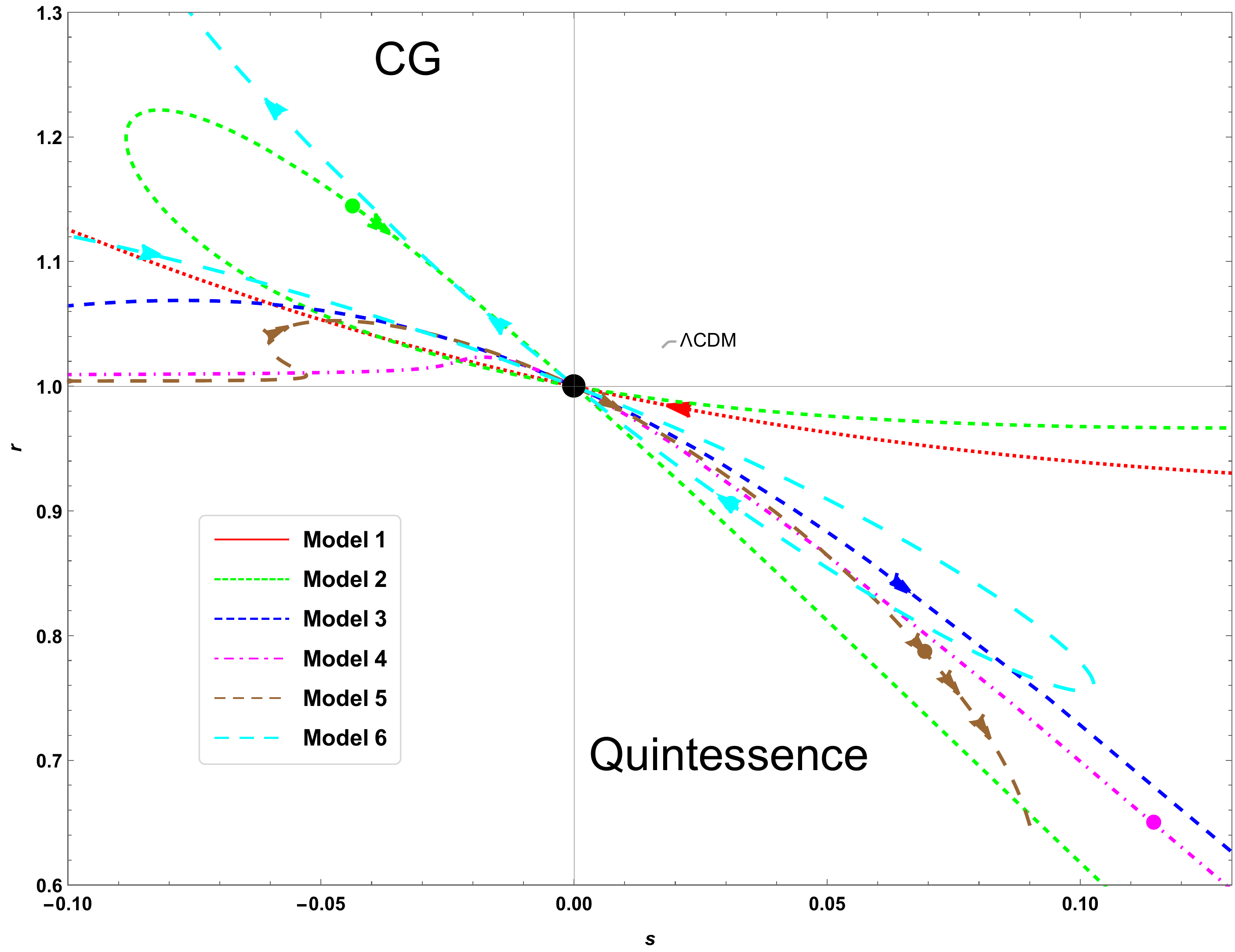}
	\caption{The local enlargement of Fig.\ref{fig:S-R}. }
	\label{fig:S-R-L}
	
\end{figure}

	According to the best-fit values given in Tabs.\ref{fit1} and \ref{fit2}, we plot the evolution of the statefinder diagnostic parameter pair $\left\{r, s\right\}$  for the six DE models in Fig.\ref{fig:S-R}, and also give the local enlargement in Fig.\ref{fig:S-R-L} for clarity. The arrows indicate the directions of evolution of the models. And, we also compare six DE models with the CG, quintessence models and $\Lambda $CDM model. We find that Model 1 behaves like quintessence model at early time, then makes transition from quintessence to $\Lambda $CDM fixed point, and finally gets into the CG region. Model 2 behaves like quintessence at early time, and passes through $\Lambda$CDM point as it evolves. After performing a swirl, it lies at quintessence region in the future. Model 5 is exactly the opposite. Its early evolution resembles CG model, then crosses the $\Lambda$CDM point and swivels round, finally enters the CG region in the future. The Evolutionary trajectories of Models 3, 4 and 5 are like  CG model at early time, then they crosses the $\Lambda$CDM fixed point into quintessence region. But the evolutionary trajectories of the two models do not overlap. Thus, the six DE models can be distinguished from each other and from  $\Lambda$CDM, CG, quintessence models by using the statefinder diagnostics method.

	\subsection{ Om diagnostics}

	Another diagnostic tool is often used to distinguish DE models and to deeply understand the constructed cosmological models, called $Om(z)$ diagnostics \cite{Sahni:2008xx,Zunckel:2008ti}. $Om(z)$ is defined as
	\begin{equation}\label{4.3}
		Om\left ( z \right ) =\frac{E^{2}\left ( z \right ) -1}{\left ( 1+z \right 	)^{3}-1 }. 
	\end{equation}
	
\begin{figure}[h]
		
		\centering
		\includegraphics[width=0.71\linewidth]{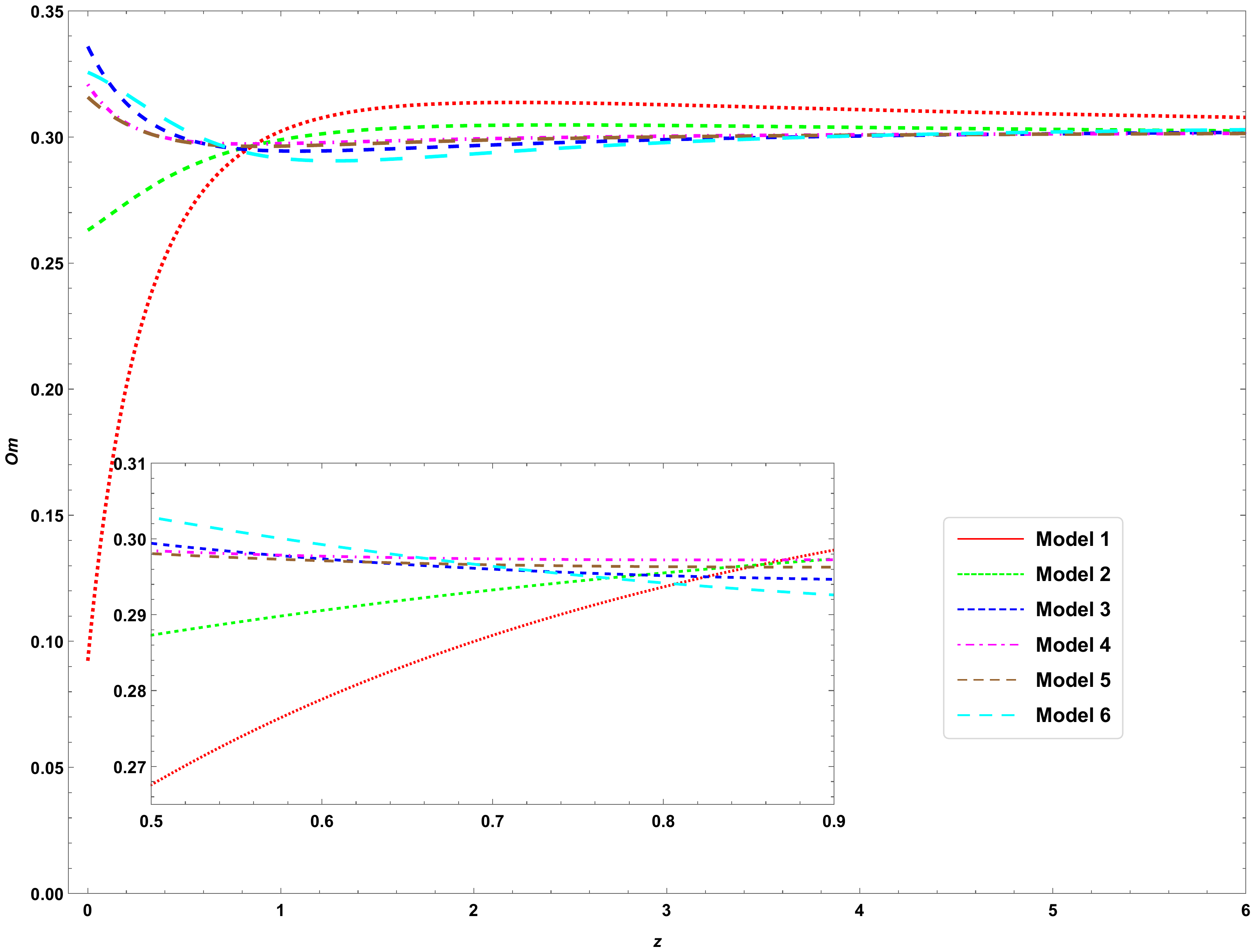}
		\caption{Evolutions of the six DE models in $Om-z $ plot. }
		\label{fig:Omz}
		
	\end{figure}

	The slope of $Om(z)$ can distinguish different DE models. The positive, negative and null slopes of $Om(z)$ correspond to phantom $\left(w<-1\right)$, quintessence $\left(w>-1\right)$ and $\Lambda $CDM $\left(w=-1\right)$ models, respectively. In Fig.\ref{fig:Omz}, we plot the evolution curves of $Om(z)$ to these six DE models by using the best-fit values given in Tabs.\ref{fit1} and \ref{fit2}. None of these models have an evolution curve with a null slope. From the evolution trends of these six models in Fig. \ref{fig:Omz}, we find that Models 2-6 are not significantly distinguishable in the early stage, but can be clearly distinguished at $z<1$. It is obvious that the results of the $Om(z)$ diagnostic are consistent with the statefinder results.

	\subsection{Growth factor}
	
	During the evolution of the universe, gravity can increase the amplitude of matter perturbations, especially at the period of matter dominance. DE not only accelerates the expansion of the universe, but also changes the growth rate of matter perturbations. On the other hand, different models possibly give the similar  accelerated expansion at late time, but they may produce different growths of matter perturbations\cite{Starobinsky:1998fr,Stephenson F R,Fry:1985zy,Silveira:1994yq,Wang:1998gt,Gong:2008fh,Gong:2009sp,Gannouji:2008jr,Nesseris:2007pa,Polarski:2016ieb}. Therefore, in addition to using observational data to distinguish DE models, exploring the effect of DE models on the growth rate of matter perturbations on large scales in the universe is another available way. Here we will analyse the types of the six DE models by considering the growth rate of matter density perturbations. Assuming that the fluid in the universe satisfies the continuity equation, Euler's equation, and Poisson's equation  as follows
	\begin{equation}\label{4.4}
		\dot{\rho }+\bigtriangledown\cdot \left ( \rho v \right ) =0,	
	\end{equation}
	\begin{equation}\label{4.5}
		\dot{v}+\left ( v\cdot \bigtriangledown  \right ) v+\frac{1}{\rho}\bigtriangledown p +\bigtriangledown \phi =0, 
	\end{equation}
	\begin{equation}\label{4.6}
		\bigtriangledown ^{2}\phi =4\pi G \rho, 
	\end{equation}
	where $\phi$ is the gravitational potential, $\rho$ is the fluid density, $p$ is the pressure, and $v$ is the velocity of fluid motion. Assume that Eqs.(\ref{4.4})-(\ref{4.6}) have perturbative solutions $\rho =\rho _{0}+\delta \rho  ,v=v_{0}+\delta v,p=p_{0}+\delta p,\phi =\phi _{0}+\delta \phi $. Introducing a new quantity $\delta\left(t,r\right)=\frac{\delta\rho}{\rho_{m}}=\delta\left(t\right){\rm exp}\left(ik\cdot x\right)$ and further assuming that the system is an adiabatic state, one can get the linear perturbation equation
	\begin{equation}\label{4.7}
		\ddot{\delta }(t)+2H\dot{\delta}(t)-4\pi G \rho _{m}\delta (t) =0. 
	\end{equation}
	From Eqs.(\ref{Fre}) and (\ref{energy}), one has the evolutionary equation of the material density parameter $\Omega_{m}=\frac{\rho_{m}}{3H^{2}}$, 
	\begin{equation}\label{4.8}
		\frac{{\rm d}\Omega_{m}}{{\rm dln}a}=3w\left(1-\Omega_{m}\right)\Omega_{m}.
	\end{equation}

	In general, the growth of perturbation is described by introducing a growth factor
	\begin{equation}\label{4.9}
		f=\frac{{\rm dln}\delta_{+}}{{\rm dln}a}.
	\end{equation}
  Where $\delta_{+}$ denotes the growth solution of the perturbation equation. Then, using Eqs.(\ref{4.8}) and (\ref{4.9}), the perturbation equation Eq.(\ref{4.7}) is rewritten as
	\begin{equation}\label{4.10}
		\frac{{\rm d}f}{{\rm dln}a}+f^{2}+\left [\frac{1}{2}-\frac{3}{2}\left ( 1-\Omega _{m} \right )w \right ]f=\frac{3}{2}\Omega_{m}.  
	\end{equation}
	
	We assume that the perturbation equation has a good approximate solution $f=\Omega_{m}^{\gamma}$\cite{Stephenson F R,Gong:2009sp,Fry:1985zy,Silveira:1994yq,Wang:1998gt,Gong:2008fh}, where $\gamma$ is the growth index. Substituting $f=\Omega_{m}^{\gamma}$ into Eq.(\ref{4.10}), one gets
	\begin{equation}\label{4.13}
		3w (1-\Omega_{m})\Omega_{m}{\rm ln}\Omega_{m}\frac{{\rm d}\gamma }{{\rm d}\Omega_{m}}-3w (\gamma -\frac{1}{2})\Omega_{m}+\Omega_{m}^{\gamma}-\frac{3}{2}\Omega_{m}^{1-\gamma}+3w \gamma-\frac{3}{2}w  +\frac{1}{2}=0.    
	\end{equation}
	Ignoring radiation, we have $\Omega_{m}+\Omega_{de}=1$. Setting a new variable $x=1-\Omega_{m}$, taking the Taylor expansion of Eq.(\ref{4.13}) near $x=0$ and letting $\gamma=\gamma_{0}+\gamma_{1}x+\gamma_{2}x^{2}+\cdot \cdot\cdot $, we can obtain the growth index\cite{Wang:1998gt,Gong:2009sp}, 
	\begin{equation}\label{4.14}
		\gamma =\frac{3(1-w )}{5-6w }-\frac{3(w -1)(3w -2)}{2(5-6w )^{2}(12w -5)}x
		-\frac{(w -1)(-194+1131w -1908w ^{2}+972w ^{3})}{4(5-6w ^{4})(12w -5)} x^{2}+ o (x^{3}).  
	\end{equation}
	
\begin{figure}[h]
		
		\centering
		\includegraphics[width=0.6\linewidth]{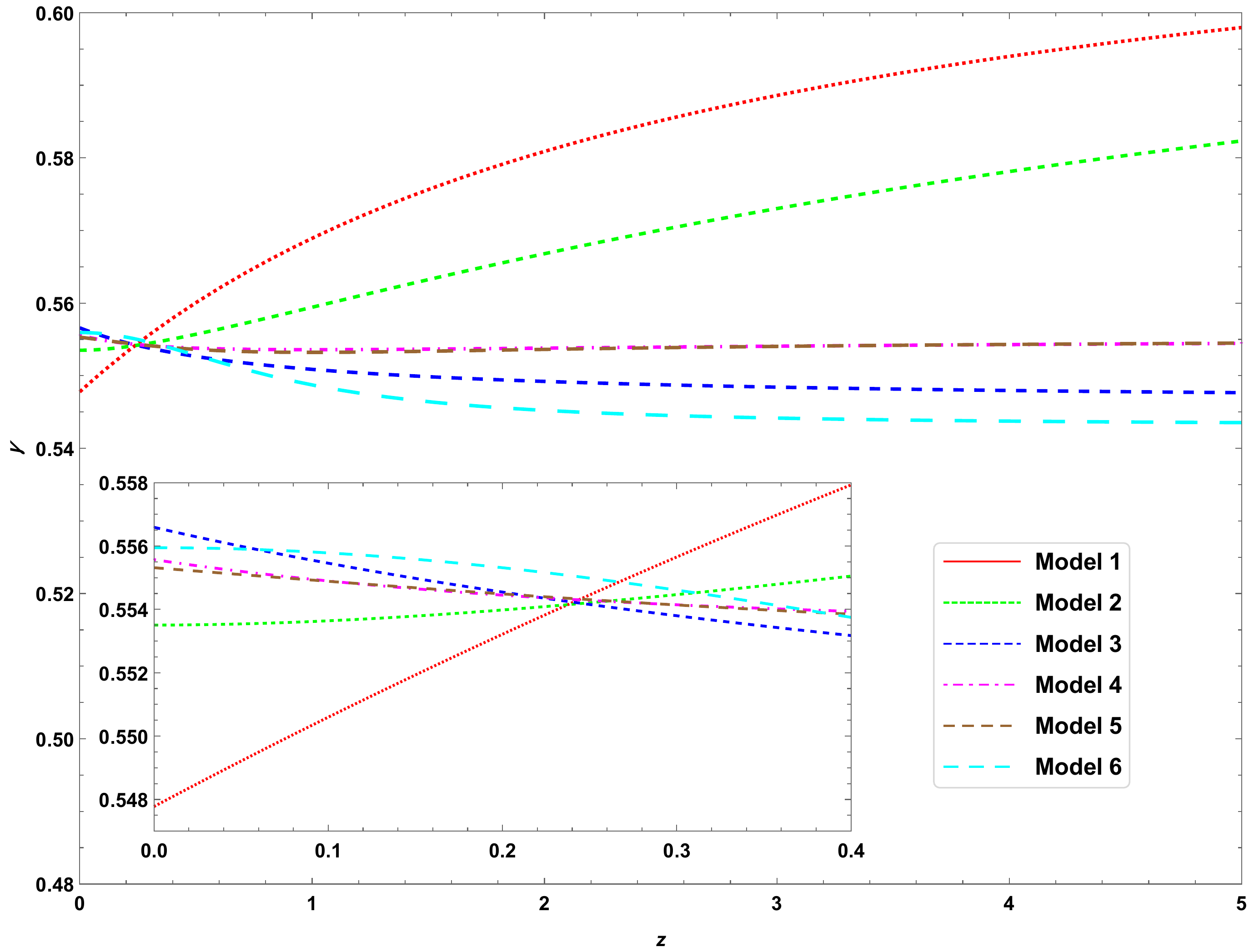}
		\caption{The evolution of growth index with redshift in different DE models.}
		\label{fig:d_z}
		
	\end{figure}

\begin{figure}[h]
		
		\centering
		\includegraphics[width=0.6\linewidth]{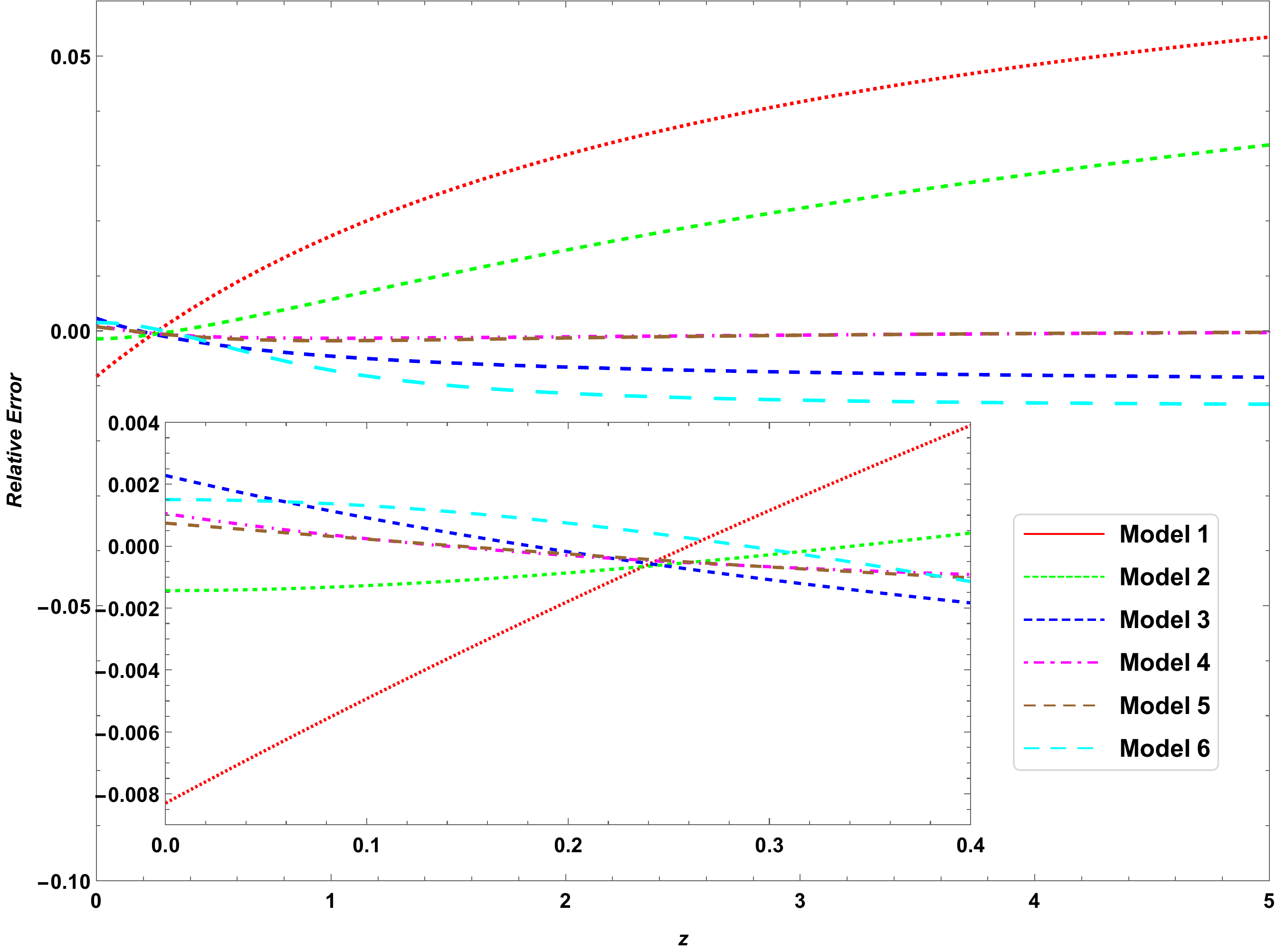}
		\caption{Evolution of relative errors of the growth factor approximation between $\Lambda $CDM model and different DE models with redshift.}
		\label{fig:c_z}
		
	\end{figure}

	From Eq.(\ref{4.14}), it is clear that the value of $\gamma$ is model-dependent. The EoS of $\Lambda $CDM model is a constant $w=-1$, then $\gamma\approx0.554698$. Based on the best-fit values in Tabs.\ref{fit1} and \ref{fit2}, we give the growth index of each model as a function of redshift, as shown in Fig.\ref{fig:d_z}. Furthermore, we plot in Fig. \ref{fig:c_z} the relative error $\sigma=\frac{f_{{\rm \Lambda CDM}}-f_{\rm Model}}{f_{\rm Model}}$ \cite{Gong:2009sp} of the growth factor approximation between the $\Lambda $CDM model and each DE models. We notice that although approximations of $f_{{\rm \Lambda CDM}}$ and $f_{\rm Model}$ are very close with relative errors in the thousands of parts, we can still distinguish the six DE models from the $\Lambda$CDM model.
	

	\section{Conclusion  and discussion}{\label{Sec.5}}
	
	In the face of many kinds of DE models, the observation data play an extremely important role to constrain the parameter space of the DE models. In this paper, we mainly use two sets of combined data: JLA+CMB+BAO+OHD (JCBH) and Pantheon+CMB+BAO+OHD (PCBH) to constrain the six parameterized DE models. The fitting results of the two combined datasets indicate that Models 3-6 are supported by recent observations, but AIC and BIC results show that $\Lambda$CDM is still the best model. So it is feasible to use the observational data to effectively distinguish and compare the DE models. Comparing the PCBH with the JCBH, we find that although covering a larger redshift range, the PCBH still has poor constraints and broadens the uncertainty of the model parameters, and is also unable to relieve the $H_{0}$ tension. The reason may be in the precision and the numbers of the observed data. As is known, large redshift data points have larger relative errors than small redshift data points. Therefore, although the PCBH data cover a larger redshift range, they give poor constraints on model parameters (see Ref.\cite{DiValentino:2020evt} for a detailed discussion). Then we expect more measurement results to help solve this problem, as mentioned in Ref.\cite{LSST:2008ijt}.

	We also use some geometrical methods to distinguish the DE models. By using statefinder and Om diagnostics, we can distinguish the six parameterized models from each other and from $\Lambda$CDM, CG, and quintessence models. The two diagnostic results are consistent. In addition, we analyse the growth factors of the matter density perturbations of the six DE models and compare them with the $\Lambda$CDM model. The results show that these DE models can be clearly distinguished. Actually, direct parameterization of growth factor or growth index can also be used to understand the properties of DE \cite{Gong:2009sp,Wu:2009zy,Chen:2009ak,Buenobelloso:2011sja,Gupta:2011kw,Mehrabi:2018dru,Velasquez-Toribio:2020xyf}, and observational constraints may further distinguish different models. In the future, this issue is worth further study by considering new forms of parameterization and using more new observational data. 
	
	The problems of $H_0$ tension and $\sigma_8$ tension are very important in modern cosmology. In this paper, we mainly focus on testing and performing constraints on some parameterized dark energy models with latest data. The $H_0$ tension does not seem to be relieved in these models. The best fitting value for $H_0$ is around 67.84-70.97 with 1$\sigma$ confidence about $\pm 1.0$. It means that one needs some mechanism to alleviate the $H_0$ tension for these models. The root mean square of the amplitude of matter perturbations\cite{Planck:2018vyg} $\sigma_8$ also has tension between the result from the low-redshift probes such as the weak gravitational lensing and galaxy clustering and the value from CMB obseravations\cite{Planck:2018vyg,Douspis:2018xlj}. A way to solve this problem is to introduce a friction between dark matter and dark energy \cite{Poulin:2022sgp} (see \cite{Escudero:2022rbq} for other possible solutions). The models discussed in this paper are lack of corresponding methods to alleviate this problem. The problems of $H_0$ tension and $\sigma_8$ tension are worthy of deep study for the parameterized dark energy models and we leave them to our next work.

\acknowledgments
This work is supported by National Science Foundation of China grant Nos.~11105091.

\end{document}